\documentclass{article}
\usepackage{srcltx}
\usepackage{amsmath}
\usepackage{amssymb}
\usepackage{amsfonts}
\usepackage{latexsym}
\usepackage{textcomp}
\usepackage{multirow}
\usepackage{booktabs}
\usepackage{url}
\usepackage{array}
\usepackage{marvosym}
\usepackage{graphics}
\usepackage{graphicx}
\usepackage{subfigure}
\usepackage{epsfig}
\usepackage[numbers,square]{natbib}

\graphicspath{{./Figures/}}

\title{An analysis of high-frequency cryptocurrencies prices dynamics using permutation-information-theory quantifiers}

\author{Aurelio F. Bariviera \\
{\tiny Universitat Rovira i Virgili,Department of Business. }\\
{\tiny Av. Universitat 1, 43204 Reus Spain}\\
{\tiny  Universidad del Pac\'ifico. Lima, Per\'u}\\
{\tiny  \texttt{aurelio.fernandez@urv.cat}}\\
Luciano Zunino \\
{\tiny  Centro de Investigaciones \'Opticas (CONICET La Plata - CIC)}\\
{\tiny C.C. 3, 1897 Gonnet, Argentina}\\
{\tiny  Departamento de Ciencias B\'asicas, Facultad de Ingenier\'ia}\\
{\tiny Universidad Nacional de La Plata (UNLP), 1900 La Plata, Argentina}\\		
{\tiny  \texttt{lucianoz@ciop.unlp.edu.ar}}\\
Osvaldo A. Rosso\\
{\tiny  Departamento de Inform\'atica en Salud, Hospital Italiano de Buenos Aires }\\
{\tiny \& CONICET, C1199ABB Ciudad Aut\'onoma de Buenos Aires, Argentina.}\\
{\tiny Instituto de F\'{\i}sica, Universidade Federal de Alagoas, }\\
{\tiny Av.\ Lourival Melo Mota, s/n, 57072-970 Macei\'o, AL, Brazil}\\
{\tiny Complex Systems Group, Facultad de Ingenier\'{\i}a y Ciencias Aplicadas,}\\
{\tiny Universidad de los Andes, Las Condes, 12455 Santiago, Chile}\\
{\tiny 	\texttt{oarosso@gmail.com}}
 }
\begin{document}
\maketitle

\begin{abstract}
This paper discusses the dynamics of intraday prices of twelve cryptocurrencies during last months' boom and bust. The importance of this study lies on the extended coverage of the cryptoworld, accounting for more than 90\% of the total daily turnover. By using the complexity-entropy causality plane, we could discriminate three different dynamics in the data set. Whereas most of the cryptocurrencies follow a similar pattern, there are two currencies (ETC and ETH) that exhibit a more persistent stochastic dynamics, and two other currencies (DASH and XEM) whose behavior is closer to a random walk. Consequently, similar financial assets, using blockchain technology, are differentiated by market participants. \\
{\bf keywords:} cryptocurrency; permutation entropy; permutation statistical complexity; complexity-entropy causality plane; informational efficiency

\end{abstract}

\section{\label{sec:Intro} Introduction}
According to the traditional definition, a currency has three main properties: (i) it serves as a medium of exchange, (ii) it is used as a unit of account and (iii) it allows to store value. Traditional currencies are issued by central banks, on behalf of nation states, and their values are related to the confidence in the central bank policies and in the economy in which such currencies are based on. A few years ago, a new type of tradable asset, called broadly cryptocurrencies, emerged. The first and most world known is Bitcoin (BTC). It was created following the publication of a manuscript written by an unknown author under the pseudonym ``Nakamoto''\cite{Nakamoto}. Contrary to standard fiat money, its creation is not linked nor endorsed by  any central bank and/or government. It is a fully private creation of virtual money, whose value is not intrinsically based on any precious metal or any other underlying asset. Consequently, its intrinsic value is zero\cite{Cheah2015}. Cryptocurrencies are based on a new technology called blockchain. Its main innovation is that transactions, instead of being validated by a central authority or clearing house, is done by several markets participants, who compete to validate them by solving complex cryptologic algorithms. In turn, the winner in this validation quest is rewarded with some amount of the cryptocurrency he/she is validating. This decentralized and encrypted transaction ledger makes, according to those who are in favor of this technology, a more reliable validation than the centralized alternative. 

The ecosystem of cryptocurrencies has been growing at an increasing pace, and now there are around 1000 active and tradable cryptocurrencies, using blockchain or similar protocols. Daily transactions are worth several millions of dollars, and in recent times a growing literature is devoted to the study of different aspects of this new asset. 

The aim of this paper is to study the informational efficiency of the twelve most important cryptocurrencies, using high-frequency data. All cryptocurrencies rely in a similar blockchain technology, making them similar from a technical point of view. However, none of them have any real or tangible asset in order to price them. Consequently, the comparative analysis aims to test if these instruments have different underlying (unobservable) dynamical structure. This article contributes to the literature in three important aspects. First, we expand the empirical studies analyzing this new asset type. Second, we compare the dynamic behavior of the twelve major cryptocurrencies. Third, we describe the temporal evolution of informational efficiency using high-frequency data. The rest of the paper is organized as follows: Section~\ref{sec:literature} describes the recent emerging literature on Bitcoin and other cryptocurrencies, Sections~\ref{sec:methodology} and \ref{sec:BP} introduce the methodology used in the paper, Section~\ref{sec:data} presents the set of data and discusses the results of our empirical analysis, and, finally, Section~\ref{sec:conclusions} draws the main conclusions.

\section{\label{sec:literature} Brief literature review}
The study of cryptocurrencies has different branches that spans law, computer science and economics. The great innovation in Nakamoto's paper\cite{Nakamoto} was probably not the creation of BTC, but the development of an open-source, decentralized online payment system. In other words, financial transactions could be done, at a very reduced fee\cite{Kim2017}, bypassing the established international banking system. Even more, due to encryption, parties are note required to disclose their true identity. This feature could arise concerns within the law community about the use of BTC for illegal purposes. Computer science literature focus its interest on the technical design of the blockchain technology, security of cryptographic protocols, vulnerabilities, energy consumption, etc. Finally, financial and monetary economics focus mainly in either the economic determinants of BTC price and in its informational efficiency. We will focus in this latter aspect.

According to the classical definition by Fama\cite{Fama1970}, a market is informationally efficient if prices convey all relevant information. In other words, and limiting the information set to the series of prices of a given asset, we say that the market for that asset is efficient if the current price incorporates the information of past prices. As a corollary of such definition, the use of past prices for future prices forecasting is futile. Samuelson\cite{Samuelson65} established that the time series of prices of any given speculative asset should behave as a random walk (RW). The empirical literature in financial economics found several deviations from the RW hypothesis. In fact, Bariviera and coauthors have shown the presence of time varying long-range dependence in the Thai Stock Market\cite{Bariviera11}, studied the effect of the 2008 financial crisis on the informational efficiency of European sovereign bonds\cite{Bariviera2013epjb}, and they have also found an asymmetric response in the stochastic characteristics of European corporate and sovereign bonds\cite{BaGuMa12}. Other authors have studied the relationship among predictability political crises and market crashes\cite{KimShamsuddinLim11}.

Regarding the cryptocurrency markets, most of the literature concentrates its efforts in the analysis of BTC. However, the cryptocurrency ecosystem is populated by hundreds of competitors to BTC. Conmarketcap\cite{coinmarketcap} gathers information of around 1000 different active currencies. In this sense, our paper gives a broader picture of this virtual market by analyzing other eleven cryptocurrencies in addition to the classical BTC.

Cheah and Fry\cite{Cheah2015} found speculative bubbles in BTC market. Urquhart\cite{Urquhart2016} reported informational inefficiency in the BTC market from 2013 until 2016. Similarly, Nadarajah and Chu\cite{Nadarajah2017} found that the time series behavior of BTC is not consistent with the Efficient Market Hypothesis (EMH), and Bariviera\cite{Bariviera20171} has shown a reduced long-term memory effect in the period 2013-2016. Finally, Bariviera~\textit{et al.}\cite{Barivieraetal2017} found that the long-term memory profile of BTC time series is similar at different time scales. It is also reported prices clustering at round numbers (with 00 decimals)\cite{Urquhart2017}.

\section{\label{sec:methodology} Information Theory quantifiers}
The departing point for many empirical studies in economics is a time series. Financial markets, and more precisely the growing cryptocurrency markets, provide abundant material to process. Taking into account that each transaction is recorded electronically, and that there are thousands of transactions per hour, the researcher can select data with different granularity. The abundance of data allows the introduction of more advanced techniques, mostly derived from econophysics, in order to shed light on economic phenomena. 

Information-theory-derived quantifiers could be very helpful to uncover information conveyed by financial time series. The use of entropy quantifiers in the financial literature can be traced back to the 1960s, with papers by Theil and Leenders\cite{TheilLeenders65}, Fama\cite{Fama65entropy}, and Dryden\cite{Dryden68}. These papers may be considered isolated examples on the use of this technique, which was only recovered in recent times, by the econophysics literature. In this line, Martina~\textit{et al.}\cite{Martina2011} and Ortiz~\textit{et al.}\cite{OrtizCruz12} applied entropy and multiscale entropy analysis to assess crude oil price efficiency. Alvarez-Ram\'irez~\textit{et al.}\cite{AlvarezRamirez2012} also used entropy methods to quantify the dynamics of the informational efficiency of the US stock market over the last 70 years. 

Shannon entropy is a very natural and common way to measure the degree of disorder in a system. According to Shannon and Weaver\cite{book:shannon1949}, given a discrete probability distribution $P=\{ p_i \in {\mathbb R};~p_i \geq 0;~i= 1,\dots,M\}$, with $\sum_{i=1}^M p_i=1$, Shannon entropy is defined as:
\begin{equation}
{\cal S}[P]= -\sum_{i=1}^M p_i \ln{p_i}.
\label{eq:entropy}
\end{equation}
 This quantifier equals zero if the patterns are fully deterministic and reaches its maximum value for a uniform distribution. 

However, analyzing time series by means of Shannon entropy alone could fall short. Feldman and Crutchfield\cite{FeldmanCrutchfield98} and Feldman~\textit{et al.}\cite{FeldmanMcTague08} advocate that an entropy measure does not quantify the degree of structure or patterns present in a process, and that a measure of statistical complexity must be introduced into the analysis in order to characterize the system's organizational properties. Mart\'in~\textit{et al.}\cite{Martin2003} and Lamberti~\textit{et al.}\cite{Lamberti2004119} have introduced a statistical complexity measure, based on the functional form developed by L\'opez-Ruiz~\textit{et al.}\cite{LMC95}, defined in the following way:
\begin{equation}
{\cal C}_{JS} [P,P_e] = {\cal H}_S [P] {\cal Q}_J[P,P_e] 
\label{eq:complexity}
\end{equation}
where ${\cal H}_S [P]=S[P]/S_{\max}$ is the normalized Shannon entropy, $P$ is the discrete probability distribution associated with the time series under analysis, $P_e$ is the uniform distribution and ${\cal Q}_J [P,P_e]$ is the so-called disequilibrium:
${\mathcal Q}_J [P,P_e] = Q_0 \{ S[ (P+P_e)/2 ] - S[P]/2 - S[P_e]/2 \}$ with $Q_0$ a normalization constant. This disequilibrium is defined in terms of the Jensen-Shannon divergence, which quantifies the difference between two probability spaces. Mart\'in~\textit{et al.}\cite{paper:martin2006} demonstrated the existence of upper and lower bounds for generalized statistical complexity measures such as ${\cal C}_{JS}$. Additionally, as highlighted in Soriano~\textit{et al.}\cite{Soriano2011a}, the statistical complexity is not a trivial function of the entropy because it is based on two probability distributions. 

The planar representation of these two quantifiers, called the complexity-entropy plane, has been introduced in the econophysics literature for characterizing the informational efficiency of several markets. For example, to rank efficiency in stock markets\cite{Zunino2010a,ZuninoCausality10}; to rank efficiency in commodity markets\cite{ZuninoPermutation11}; to link informational efficiency with sovereign bond ratings\cite{Zunino2012}; to assess the impact of the establishment of a common currency and a deep and wide financial crisis in European sovereign bonds time series\cite{Bariviera2013epjb}; and to detect Libor manipulation\cite{EPJB2015,BarivieraRSTA2015}.

\section{\label{sec:BP} Bandt-Pompe time series symbolic encoding}
Many economic phenomena produce observable magnitudes, which are registered at evenly distributed times. These observations, \textit{i.e.} time series, are the raw materials used by quantitative analysts to model and scrutinize complex phenomena. This research area is broadly known as time series analysis. One of its goals is to describe the nature of the generating process. We can safely assume that a straight departing point for this task is to find the appropriate probability density function (PDF) associated with the time series. There are several competing methodologies for PDF estimation. Beyond traditional histogram technique, and without attempting to be exhaustive, we can cite: binary symbolic dynamics\cite{Mischaikow1999}, Fourier analysis\cite{PowellPercival}, wavelet transform\cite{Rosso2001Wavelet}, and ordinal patterns\cite{BandtPompe02}. The suitability of each method depends on the very own characteristics of the data. The methods for symbolic analysis of time series discretize raw series and transform it into a series of symbols. These kind of methods are very powerful because they are rarely affected by the presence of observational noise\cite{FinnGoettee2003}. This property is specially important in the analysis of economic time series, where noise is a traditional feature. Among the symbolic-based techniques for PDF estimation, the Bandt and Pompe (BP) methodology\cite{BandtPompe02} has the advantage of considering time causality in its estimation. This symbolic methodology is robust to the presence of (observational) noise and requires no \textit{a priori} model assumption, except weak stationarity. The starting point of this method is to consider the ordinal structure of $D-$dimensional partitions of the time series. ``Partitions'' are devised by comparing the order of neighboring relative values rather than by apportioning amplitudes according to different levels.

Let consider a time series ${\mathcal S}(t)=\{x_t; t = 1,\dots,N\}$, an embedding dimension (pattern length) $D>1$ $(D \in {\mathbb N})$, and an embedding delay (sampling frequency) $\tau$ $(\tau \in {\mathbb N})$, the BP-pattern of order $D$ generated by
\begin{equation}
\label{eq:vectores}
s \mapsto \left(x_{s-(D-1)\tau},x_{s-(D-2)\tau},\dots,x_{s-\tau},x_{s}\right),
\end{equation}
is the one to be considered. To each time $s$, BP method assigns a $D$-dimensional vector that results from the evaluation of the time series at times $s-(D-1)\tau,s-(D-2)\tau,\dots,s-\tau$, and $s$. Clearly, the higher value of $D$, the more ``time causality'' is incorporated into the ensuing vectors. By the ordinal pattern of order $D$ related to the time $s$, BP mean the permutation $\pi = (r_0,r_1,\dots,r_{D-1})$ of $(0,1,\dots,D-1)$ defined by
\begin{equation}
\label{eq:permuta}
x_{s-r_{D-1}\tau} \le x_{s-r_{D-2} \tau} \le \dots \le x_{s-r_{1} \tau}\le x_{s-r_0 \tau}.
\end{equation}
In this way the vector defined by Eq. (\ref{eq:vectores}) is converted into a definite symbol $\pi$. So as to get a unique result, BP consider that $r_i<r_{i-1}$ if $x_{s-r_{i} \tau}=x_{s-r_{i-1} \tau}$. This is justified if the values of ${x_t}$ have a continuous distribution so that equal values are very unusual.

For all the $D!$ possible orderings (permutations) $\pi_i$ when embedding dimension is $D$, their associated relative frequencies can be naturally computed according to the number of times this particular order sequence is found in the time series, divided by the total number of sequences,
\begin{equation}
\label{eq:frequ}
p(\pi_i)= \frac{\sharp \{s|s \leq N-(D-1)\tau ; (s) \quad \texttt{has type}~\pi_i \}}{N-(D-1)\tau}.
\end{equation}
In the last expression, the symbol $\sharp$ stands for ``number". Thus, an ordinal pattern probability distribution $P=\{p(\pi_i),i=1,\dots,D!\}$ is obtained from the time series.

The ordinal pattern PDF is invariant with respect to nonlinear monotonous transformations. Accordingly, nonlinear drifts or scalings artificially introduced by a measurement device will not modify the quantifiers' estimation, a nice property if one deals with experimental data (see, \textit{e.g.}, Saco~\textit{et al.}\cite{Saco2010}). These advantages make the BP approach more convenient than conventional methods based on range partitioning. Additional advantages of the method reside in its simplicity (we need few parameters: the pattern length/embedding dimension $D$ and the embedding delay $\tau$) and the extremely fast nature of the pertinent calculation-process\cite{paper:keller2005}. The BP methodology can be applied not only to time series representative of low dimensional dynamical systems but also to any type of time series (regular, chaotic, noisy, or reality based)\cite{BandtPompe02}. In fact, the existence of an attractor in the $D$-dimensional phase space is not assumed. The only condition for the applicability of the BP method is a very weak stationary assumption: for $k \leq D$, the probability for $x_t < x_{t+k}$ should not depend on $t$. For review of BP's methodology and its multidisciplinary applications, please see Zanin~\textit{et al.}\cite{Zanin2012} and references therein.

In this work, the normalized Shannon entropy ${\cal H}_S$ and the statistical complexity measures ${\cal C}_{JS}$ (Eq.~(\ref{eq:complexity})), are estimated using the ordinal pattern probability distribution $P=\{p(\pi_i),i=1,\dots,D!\}$. Defined in this way, these quantifiers are usually known as permutation entropy and permutation statistical complexity. They characterize the diversity and correlational structure, respectively, of the orderings present in the complex time series. The complexity-entropy causality plane (CECP) is defined as the two-dimensional (2D) diagram obtained by plotting permutation statistical complexity (vertical axis) versus permutation entropy (horizontal axis) for a given system\cite{Rosso07}. The term causality remembers the fact that temporal correlations between successive samples are taken into account through the BP recipe used to estimate both information-theory quantifiers.

\section{\label{sec:data} Data and results}
We use high-frequency price indices developed by MV Index Solutions (MVIS\textregistered). Data were obtained from Thomson Reuters Eikon terminal from one of the authors' university. Data consist of 16,031 observations of price indices, for each of the twelve cryptocurrencies detailed in Table~\ref{tab:data}. Data are equally spaced in time, being 5 minutes the time frame between each observation. The period under study spans from December 3, 2017 until February 14, 2018. This period is very interesting since cryptocurrencies exhibited an unprecedented rise and subsequent crash in their values. Consequently, it could be suitable to study the co-movement of different currencies for testing if the underlying dynamics of the different time series were the same.

\begin{table}
\caption{Data}
\begin{tabular}{lll}
Cryptocurrency & Acronym & Reuters Instrument Code (RIC) \\
\hline
Bitcoin Cash & BCH & .MVBCH \\
Bitcoin & BTC & .MVBTC \\
Dash & DASH & .MVDASH \\
Ethereum Classic & ETC & .MVETC \\
Ethereum & ETH & .MVETH \\
IOTA & IOT & .MVIOT \\
LiteCoin & LTC & .MVLTC \\
NEO & NEO & .MVNEO \\
NEM & XEM & .MVXEM \\
Monero & XMR & .MVXMR \\
Ripple & XRP & .MVXRP \\
Zcash & ZEC & .MVZEC \\
\end{tabular}
\label{tab:data}
\end{table}

In spite of the fact that Bitcoin is, undoubtedly, the most famous cryptocurrency, there are several hundreds of tradable instruments using a similar blockchain technology. As can be seen in Table~\ref{tab:capitalization}, the market is very concentrated. Our twelve selected cryptocurrencies account for 88\% of total market capitalization and 91\% of 24 hours traded volume, among the 897 ones detailed in the website \texttt{https://coinmarketcap.com/coins/views/all/}\footnote{Accessed on 14/02/2018}. Consequently, our study covers most of the cryptocurrency market.

\begin{table}
\caption{Market capitalization and 24-hour trading volume of the selected cryptocurrencies. Percentages represent the proportion of capitalization or traded volume with respect to 897 cryptocurrencies. Own elaboration based on data from \texttt{https://coinmarketcap.com/coins/views/all/}.}
\begin{tabular}{lrrrr}
& \multicolumn{2}{c}{Market Capitalization} & \multicolumn{2}{c}{Daily traded volume} \\
Acronym & \multicolumn{1}{r}{USD} & \multicolumn{1}{r}{\% of} & \multicolumn{1}{r}{USD} & \multicolumn{1}{r}{\% of} \\
& \multicolumn{1}{r}{millions} & \multicolumn{1}{r}{cryptos} & \multicolumn{1}{r}{millions} & \multicolumn{1}{r}{cryptos} \\
\hline
BCH & 22,931 & 5.5\% & 678 & 3.3\% \\
BTC & 165,007 & 39.4\% & 9,128 & 44.0\% \\
DASH & 5,355 & 1.3\% & 151 & 0.7\% \\
ETC & 3,384 & 0.8\% & 765 & 3.7\% \\
ETH & 90,727 & 21.7\% & 3,143 & 15.2\% \\
IOT & 5,698 & 1.4\% & 68 & 0.3\% \\
LTC & 12,580 & 3.0\% & 2,731 & 13.2\% \\
NEO & 7,913 & 1.9\% & 265 & 1.3\% \\
XEM & 5,049 & 1.2\% & 79 & 0.4\% \\
XMR & 4,356 & 1.0\% & 123 & 0.6\% \\
XRP & 44,039 & 10.5\% & 1,702 & 8.2\% \\
ZEC & 1,566 & 0.4\% & 104 & 0.5\% \\
& & & & \\
Total & 368,606 & 88.1\% & 18,937 & 91.3\% \\
\end{tabular}%
\label{tab:capitalization}%
\end{table}%

\begin{figure}
\centering
\includegraphics[width=0.3\columnwidth,angle=-90]{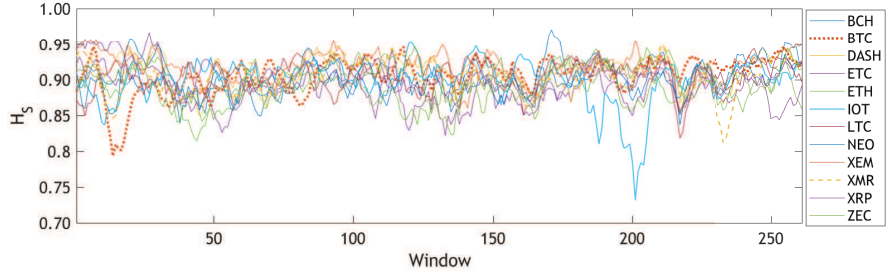}
\caption{Permutation entropy evolution associated with the selected cryptocurrencies during the observation period. Estimations were obtained by implementing sliding windows with the following parameters: $D=4$, $\tau=1$, $N=360$ and $\delta=60$.}
\label{fig:entropyevolution}
\end{figure}

One feature of this market is that its dynamics is very similar for all the assets under study. Figure~\ref{fig:entropyevolution} shows how the permutation entropy varies across time. Sliding windows of size $N=360$ data points and step $\delta=60$ have been implemented for the dynamical analysis. Behaviors are very similar for all cryptocurrencies. This could reflect coherent dynamics of the different time series.

\begin{figure}
\centering
\includegraphics[width=0.7\columnwidth]{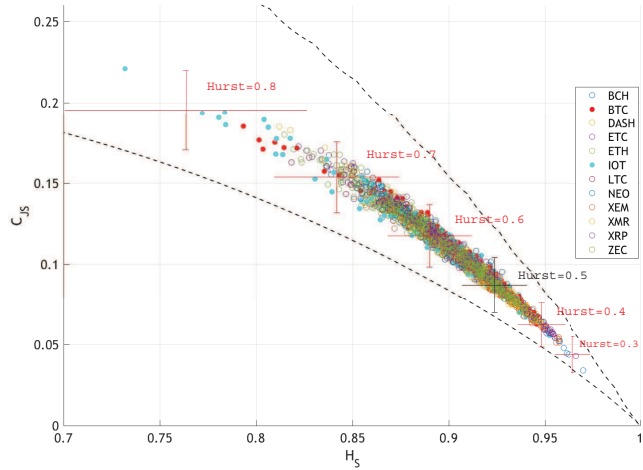}
\caption{Location of the cryptocurrencies in the CECP computed using sliding windows with the following parameters: $D=4$, $\tau=1$, $N=360$ and $\delta=60$. Black and red crosses are mean and standard deviation of 500 fractional Brownian motion (fBm) simulations of 360 data points for the Hurst exponents indicated in the figure. Dashed lines represent the upper and lower bounds of the quantifiers as computed by Mart\'in~\textit{et al.}\cite{paper:martin2006}.}
\label{fig:CECP}
\end{figure}

We can observe in Figure~\ref{fig:CECP} that time series mostly exhibit persistent behavior, reflected in a location in the CECP compatible with fractional Brownian motions (fBm) with Hurst exponents between 0.5 and 0.7. Previous studies on BTC time series reported an enhanced informational efficiency in the period 2014-2016. Nevertheless, it seems that strong bull and bear markets could lead to more coordinated movements that reduce the informational efficiency.

\begin{figure}
\centering
\includegraphics[width=0.7\columnwidth]{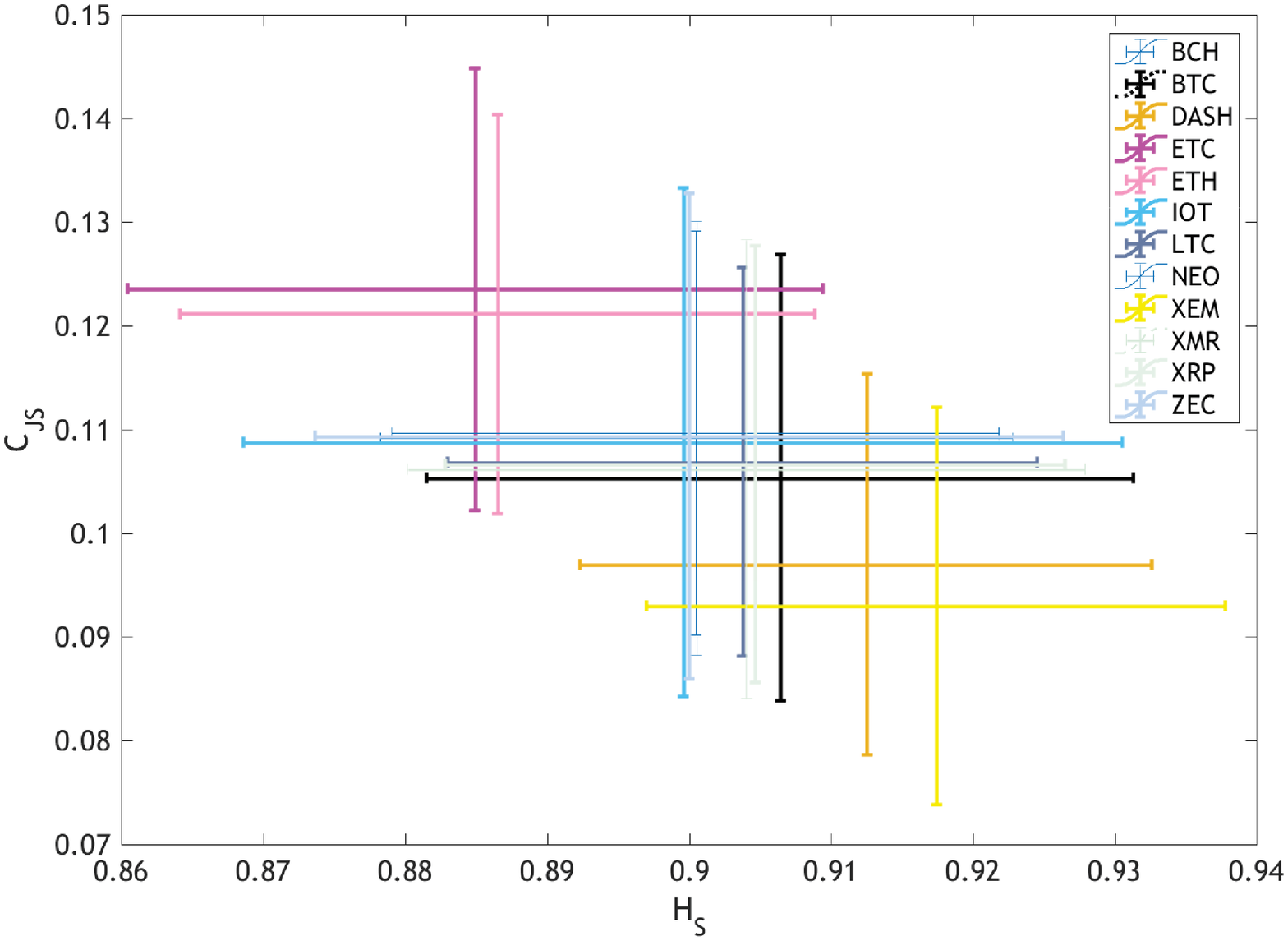}
\caption{Mean and standard deviation of each cryptocurrency in the CECP during the observation period. Quantifiers were calculated by implementing sliding windows with the following parameters: $D=4$, $\tau=1$, $N=360$ and $\delta=60$.}
\label{fig:CECPmeans}
\end{figure}

In order to verify if all cryptocurrencies follow the same stochastic process, we compute the sample mean and standard deviation of the information-theory quantifiers for each currency. We depict results in Figure~\ref{fig:CECPmeans}. We observe that BTC occupies a central position among the other currencies. Additionally, there are some other currencies more and less efficient than BTC.

Taking into account that in our framework, informational efficiency is maximal as ${\cal H}_S[P]$ approaches 1 and  ${\cal C}_{JS}[P]$ approaches 0, we compute the Euclidean distance of the mean permutation entropy and permutation statistical complexity of each currency to $({\cal H},{\cal C})=(1,0)$, as a proxy for an informational efficiency ranking. Results are displayed in Table \ref{tab:ranking}.

\begin{table}[htbp]
\caption{Informational efficiency ranking.}
\begin{tabular}{rrr}
Ranking   & Cryptocurrency &  efficiency measure \\ 
position & &  $d[({\cal H},{\cal C})-(1,0)]$ \\
 \hline 
7 &  BCH   &   0.1477 \\
3 &  BTC & 0.1409 \\
2  & DASH & 0.1306 \\
12  &ETC & 0.1688 \\
11   &  ETH & 0.1660 \\
8  & IOT & 0.1480 \\
6  & LTC & 0.1438 \\
9   &  NEO & 0.1481 \\
1  & XEM  & 0.1244 \\
5  & XMR	& 0.1431 \\
4   &  XRP	 & 0.1431 \\
10  & ZEC   & 0.1482  \\
\end{tabular}%
\label{tab:ranking}%
\end{table}%

One important finding of this paper is that informational efficiency is not related to currency size. In fact, BTC, by far the largest cryptocurrency in terms of capitalization and daily turnover, is not the most efficient one. Additionally, we compute Spearman's rho, a non parametric correlation measure, between our efficiency measure displayed in Table \ref{tab:ranking} and market capitalization and daily turnover informed in Table \ref{tab:capitalization}. The Spearman's rho of the efficiency measure against market capitalization is 0.1748 (p-value 0.5868), and against daily traded volume is 0.1225 (p-value 0.7042). In both cases the association is not statistically significant.

\begin{figure}
\centering
\includegraphics[width=0.7\columnwidth]{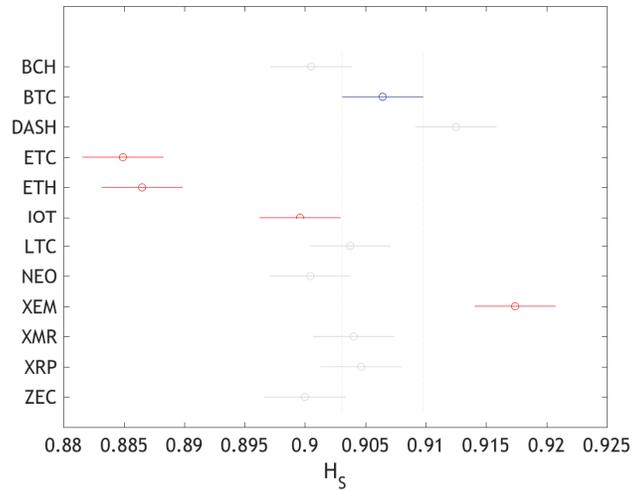}
\caption{Anova analysis. Difference of mean permutation entropy for each cryptocurrency with respect to BTC. Red lines indicate currencies whose mean permutation entropy is different from BTC (at 1\% significance for ETC, ETH and XEM, and 5\% level for IOT).}
\label{fig:anova-entropy}
\end{figure}

\begin{figure}
\centering
\includegraphics[width=0.98\columnwidth]{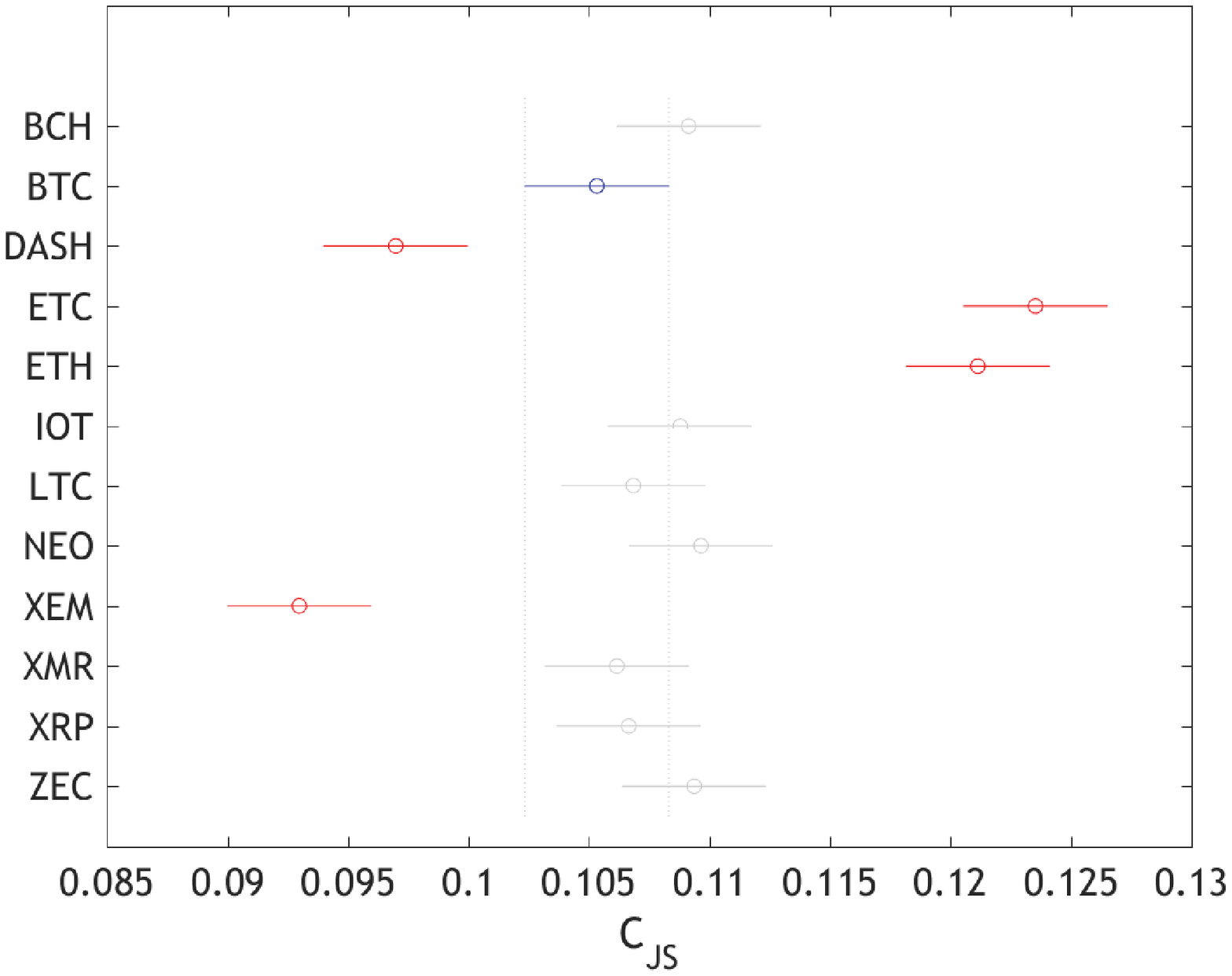}
\caption{Anova analysis. Difference of mean statistical complexity for each cryptocurrency with respect to BTC. Red lines indicate currencies whose mean statistical complexity is different from BTC (at 1\% significance level).}
\label{fig:anova-complexity}
\end{figure}

We also test using ANOVA if the mean permutation entropy and mean permutation statistical complexity are equal for all cryptocurrencies. Results are displayed in Table~\ref{tab:anova}, and we cannot accept the null hypothesis of equal mean values for either of the quantifiers among cryptocurrencies. Second, we perform ANOVA analysis for each currency \textit{vis-\`a-vis} BTC. Results are displayed in Figures~\ref{fig:anova-entropy} and \ref{fig:anova-complexity}. We observe that there are seven cryptocurrencies (displayed in light gray in the figures), whose mean entropic and complexity behavior is indistinguishable form BTC (displayed in blue in the figures). However, we reject the null hypothesis of equal mean permutation entropy of BTC, with respect to ETC, ETH, IOT and XEM (displayed in red in Figure~\ref{fig:anova-entropy}). We also reject the null hypothesis of equal permutation statistical complexity of BTC, with respect to DASH, XEM, ETC and ETH (displayed in red in Figure~\ref{fig:anova-complexity}). If we analyze these results together with the graphical representation of mean values of Figure~\ref{fig:CECPmeans}, we conclude that ETC and ETH are less efficient (more persistent) while DASH and XEM are more efficient than BTC. Actually, DASH and XEM dynamics are closer to a random walk behavior.
 One of the reasons for such behavior of ETC and ETH, could be found in the fact that this cryptocurrencies were not created with the aim of substituting paypal-like systems. Ethereum's goal is using a blockchain  for ``smart contracts'', i.e. to replace internet third parties in order to validate trusted operations \cite{Ethereum}. 

Additionally, XEM and DASH appear as the most efficient cryptocurrencies. In this case, the reason could be found in the validation design. Both currencies introduced different ways of validating blocks. XEM introduced a proof-of-importance (POI) algorithm, and an Eigentrust++ reputation system in order to check operations. Unlike BTC, DASH is comprised of three types of 'levels', with specific roles and responsibilities on the network. In addition, from the beginning the evolution, changes or upgrades in the currency can be proposed by anyone, establishing  a decentralized governance by blockchain. This situation could generate fairer transactions, which leads to a more efficient market.

\begin{table}[htbp]
\caption{Anova analysis to test the equality of means among all cryptocurrencies.}
\begin{tabular}{rrrrrr}
\multicolumn{6}{l}{ANOVA on permutation entropy} \\
Source & SS & df & MS & F & Prob$>$F \\
\hline 
Currencies & 0.2421 & 11 & 0.0220 & 39.8496 & 3.19E-81 \\
Error & 1.7231 & 3120 & 0.0006 & & \\
Total & 1.9651 & 3131 & & & \\
& & & & & \\
\hline
\multicolumn{6}{l}{ANOVA on statistical complexity} \\
Source & SS & df & MS & F & Prob$>$F \\
\hline
Currencies & 0.2039 & 11 & 0.0185 & 42.3817 & 2.12E-86 \\
Error & 1.3647 & 3120 & 0.0004 & & \\
Total & 1.5686 & 3131 & & & \\
\end{tabular}%
\label{tab:anova}%
\end{table}%

\section{\label{sec:conclusions} Conclusions}
We studied high-frequency data of the cryptocurrency market during a very special period of boom and bust. Our paper reports detailed behaviors of the twelve most important cryptocurrencies, which cover 88\% of market capitalization and over 91\% of daily turnover. We detect that the majority of the currencies exhibit a similar behavior, compatible with some kind of persistent stochastic dynamics with Hurst exponents between 0.5 and 0.7. However, we can identify four cryptocurrencies whose behaviors are different from the rest. ETC and ETH exhibit more persistent behavior than the others, reflected in smaller mean permutation entropies and larger mean statistical complexities. On the contrary, DASH and XEM average behaviors are closer to a random walk. Our results uncover that, inside the cryptocurrency ecosystem, distinct behaviors emerge. Even though the majority of the market follow the behavior of the leader (BTC), some alternative cryptocurrencies follow differentiated dynamics, which could indicate that these assets are not as homogeneous as expected.  The reason for such behavior could be found in the special characteristics of these currencies. Unlike BTC, the aim of ETC and ETH is to be a vehicle for ``smart contracts'' rather than a virtual currency system. Regarding DASH and XEM, they introduced some innovations in the blockchain ecosystem and, consequently, investors could see them as more reliable assets.

\end{document}